\documentclass[11pt,twoside]{article}

\usepackage{asp2006}
\usepackage{graphicx}

\begin{document}

\title{The Eccentric Disc Instability: Dependency on Background Stellar Cluster} 
\author{Ann-Marie Madigan}

\affil{Leiden Observatory, Leiden University, P.O. Box 9513, NL-2300 RA Leiden}

\begin{abstract} 

In this paper we revisit the ``eccentric disc instability", an instability which occurs in coherently eccentric discs of stars orbiting massive black holes (MBHs) embedded in stellar clusters, which results in stars achieving either very high or low eccentricities. The preference for stars to attain higher or lower eccentricities depends significantly on the density distribution of the surrounding stellar cluster. Here we discuss its mechanism and the implications for the Galactic Centre, home to at least one circum-MBH stellar disc.

\end{abstract}

\section{Introduction}\label{sec:intro}

\indent The Galactic Centre (GC) hosts at least one clockwise rotating disc of O and Wolf-Rayet stars, most likely a result of the gravitational fragmentation of a gaseous accretion disc \citep{Lev03}. These stars are located at projected distances of $0.05 - 0.5$ pc, and have an average age of $\sim 6$ Myr \citep{Gen03a, Pau06,Lu09,Bar09}. Recent hydrodynamical simulations suggest that young stars forming in such discs can retain similar initial orbital eccentricities, which are often non-zero \citep[e.g.][]{BoR08}. The dynamics of stars in such eccentric discs can be very different from circular structures. In particular there exists an instability \citep{Mad09} that occurs in coherently eccentric discs which are embedded in nuclear stellar clusters such as the one in our GC (see review by Sch\"odel in this volume). This instability can propel stars to very high, or low, orbital eccentricities within $\sim 1$ Myr, overcoming the long relaxational time scales associated with galactic nuclei. 

The background stellar cluster (or cusp) is an integral part of the instability. In this paper we identify the reasons for this, vary the density profile of the cluster to see the effect it has on the instability, and address the possible implications of the observed flattened density profile of the cusp of late-type stars in the GC \citep{Buc09, Do09, Bar09}.

\section{The Eccentric Disc Instability}\label{sec:EDI}

To describe the physics of the instability we work with a simplified template of a galactic nucleus. We consider a disc of stars with mass $M_{\rm disc}$ orbiting a MBH of mass $M_{\bullet}$. Both are embedded in a power-law stellar cusp of mass $M_{\rm cusp}$. 

We make the assumptions that (1) $M_{\rm disc}\ll M_{\rm cusp}$, such that the apisidal precession of stellar orbits within the disc is driven by the cusp, and (2) initially each stellar eccentricity vector (which points to periapse) is aligned and similar in magnitude to those of the other stars in the disc. The stellar orbits precess with retrograde motion due to the presence of the cusp, i.e., in the direction opposite to the orbital rotation of the stars. The precession time scales as
\begin{equation} \label{eqn:t_prec}
t_{\rm prec} \sim \frac{M_{\bullet}} {N( < a) m} P(a) f(e) \quad \propto a^{\gamma-3/2} f(e) ,
\end{equation}
\noindent where $P(a) = 2 \pi \sqrt{a^3/G M_{\bullet}}$ is the period of a star with
semi-major axis $a$, $N(<a)$ is the number of stars in the cusp within $a$, $m$ is the individual mass of the stars, and $\gamma$ is the power-law index for the space-density profile of the cusp $\rho(r) \propto r^{-\gamma}$. For simplicity let us assume a power-law index $\gamma = 3/2$, such that $t_{\rm prec}$ is constant for all values of $a$. The key element of the instability is that $f(e)$ in Equation (\ref{eqn:t_prec}) is an increasing function of eccentricity, $df/de > 0$.

\begin{figure}[!ht]
	\begin{center}
	\includegraphics[scale = 0.55]{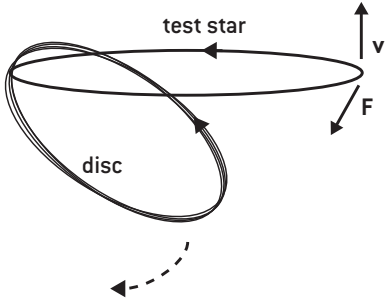}
		\caption{This schematic shows the stellar disc from above its orbital plane. The stars move counterclockwise on their orbits and precess in the opposite direction (dashed arrow). A ``test star" is also shown, with a more eccentric orbit, and lagging behind in precession. The direction of its velocity vector, ${\bf v}$, at apoapse is opposite to the coherent gravitational force, ${\bf F}$, it feels from the rest of the stars in the disc. This causes the torque, ${\bf \tau} = {\bf r} \times {\bf F}$, acting on the star from the disc to decrease its angular momentum, ${\bf J} = {\bf r} \times {\bf v}$.
	\label{f:torque_diagram}}
		\end{center}
\end{figure}

The instability works as follows: A star that has a slightly more eccentric orbit than the average star in the disc consequently has a larger $t_{\rm prec}$ and lags behind in precession; see Figure (\ref{f:torque_diagram}). Such a star feels a strong, coherent torque from the other stars in the disc, in the {\it opposite} direction of its angular momentum vector, $J = \sqrt{G M_{\bullet} a (1 - e^2)}$. As a result, its angular momentum decreases in magnitude, which is equivalent to saying that its eccentricity increases\footnote{Now, with an even higher eccentricity, the stellar orbit lags even further behind the bulk of the disc and the cycle repeats. We emphasize that the precession due to the cusp is a key element of this instability. Without it, stars in the eccentric disc would experience torques but in both directions and it would not be an unstable configuration.}. In this way, very high eccentricities can be achieved. Conversely, if a stellar orbit is slightly less eccentric than the average, it has a smaller $t_{\rm prec}$ and moves ahead in precession, thus experiencing a torque which decreases its eccentricity further below the average. 

\subsection{Dependency on Background Stellar Cluster}

\begin{figure}[!ht]
	\begin{center}
	\includegraphics[scale = 0.44, angle = 270]{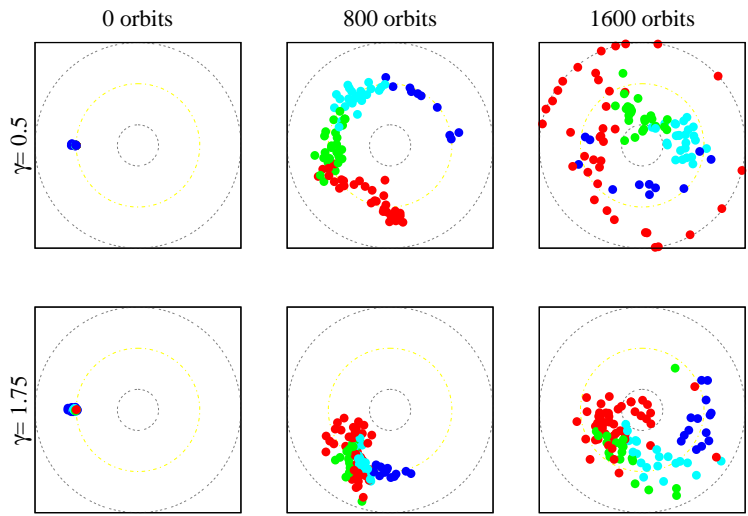}
		\caption{Evolution of the (x and y components of the) eccentricity vectors (EVs) of stars within an eccentric disc embedded in a background cusp with index $\gamma = 0.5$ (top) and $\gamma = 1.75$ (bottom). Time is in units of the initial period of the innermost orbit at $a = 0.05 {\rm pc}$; final time corresponds to $\sim \!0.8$ Myr. The innermost circle shows the initial value of the eccentricities $e = 0.6$;  the outermost is $e = 1$. The colour of the EVs indicate the semi major axes of the stars in the disc, going from red ($\sim0.05 - 0.1$ pc) outwards to $0.5$ pc (green, cyan and dark blue).	The EVs spread out as the stars complete more and more orbits, and precess with retrograde motion (clockwise here). Many more high eccentricity stars are produced in simulations with a flatter cusp ($\gamma = 0.5$) as the innermost stars lag behind in precession. In simulations with a steep cusp  ($\gamma = 1.75$), the stars precess more rapidly ($t_{\rm prec} \sim 900$ orbits) and the innermost orbits have lower average eccentricities.
	\label{f:gamma}}
		\end{center}
\end{figure}

The stellar orbits at the innermost radii of the disc (i.e, with the smallest $a$) undergo the greatest fractional change in $J$ and hence are most likely to be pushed to extreme eccentricities. Consequently, the precession rates of these orbits are important. 
Recalling that $t_{\rm prec} \propto a^{\gamma-3/2} f(e)$, we perform $N$-body simulations of an eccentric stellar disc to examine the effect of varying the index of the power-law slope of the cusp, $\gamma$, on the instability; see Figure (\ref{f:gamma}) for examples where $\gamma = 0.5$ (shallow cusp) and $\gamma = 1.75$ (steep cusp). The disc has a surface density profile $\Sigma \propto r^{-2}$, $M_{\rm disc} = 10^4 M_{\odot}$, a semi major axis distribution of $0.05 \leq a \leq 0.5$ pc and a smooth stellar cusp with $M_{\rm cusp}(<1 {\rm pc}) = 0.5 \times 10^6 M_{\odot}$ \citep[see][for details]{Mad09}. For $\gamma < 3/2$, due to the shallowness of the cusp, the innermost orbits lag behind in precession and are predisposed to being torqued to high $e$ orbits; the instability is not as effective at generating low $e$ orbits. Conversely, an increase in $\gamma$ $(> 3/2)$ results in the innermost orbits precessing faster than the bulk of the stars and hence suppresses the generation of the highest $e$ orbits.
 
\subsection{Implications for the Galactic Centre}  
A (so far unknown) percentage of stars in a disc should be formed in binary systems. These binaries can be propelled to high eccentricities by the eccentric disc instability, pass within the tidal radius of the MBH and be disrupted \citep{Hil88, YuQ03}. In the GC, this mechanism may be responsible for producing both hypervelocity stars \citep[HVS; see][and references therein]{Bro09} and the S-stars \citep{Sch02, Ghe05, Eis05, Gil09}. There are three lines of evidence to support this idea. (1) The cusp of late-type stars in the GC is observed to have a flat surface density profile \citep{Buc09, Do09, Bar09} which, if it traces the underlying mass distribution, is maximally effective in producing high eccentricity orbits with the instability. (2) \citet{Lu09} find that the observed HVS are consistent with two planar structures, one alined with the inner edge of the clockwise disc\footnote{The instability preferentially propels stars at the inner edge of the disc to extreme eccentricities. These stars are likely to have significant inclinations however which suggests that, in this scenario, the HVS will not be tightly confined to a plane.}. The HVS travel time from the GC to their current positions is too large (100 - 200 Myr) for them to have originated in the observed young disc but an older disc could potentially explain this feature. (3) \citet{Bar09} find a bimodal distribution of eccentricities in the young disc. The eccentric disc instability naturally produces a double-peaked eccentricity profile as the stars are pushed away from the average value. However, we must emphasize that a direct comparison to the observations is not possible at this time as the most eccentric stars in our simulations are the most highly inclined and would not be observed along the plane of the disc. \\

\acknowledgements
AM thanks Olivera Raki\'c for her helpful comments.

\end{document}